\documentclass[runningheads]{llncs}
\usepackage{graphicx}
\usepackage{multirow,multicol} 
\usepackage{listings} 
\usepackage{algorithm} 
\usepackage{algorithmicx,algpseudocode} 
\usepackage{array} 
\usepackage{bm}
\usepackage{amsfonts} 
\usepackage{url}
\usepackage{epstopdf}
\usepackage{booktabs} 
\usepackage[flushleft]{threeparttable} 
\usepackage{makecell} 
\usepackage{amsmath} 
\usepackage{mathtools} 
\usepackage{longtable,rotating}
\usepackage{threeparttable}
    
\usepackage{color}

\begin{document}
\title{Siamese Encoder-based Spatial-Temporal Mixer for Growth Trend Prediction of Lung Nodules on CT Scans}
\titlerunning{Spatial-Temporal Mixer}
%
\author{Jiansheng Fang \inst{1,2,3} \and Jingwen Wang \inst{2} \and Anwei Li \inst{2} \and Yuguang Yan \inst{4} \and Yonghe Hou \inst{5} \and Chao Song \inst{5} \and Hongbo Liu \inst{2} \and Jiang Liu \inst{3} \thanks{Corresponding author. Co-first authors: Jiansheng Fang, Jingwen Wang, Anwei Li. This work was supported in part by Guangdong Provincial Department of Education (2020ZDZX3043), and Shenzhen Natural Science Fund (JCYJ20200109140820699 and the Stable Support Plan Program 20200925174052004).}}


\authorrunning{Fang et al.}
%
\institute{
School of computer science and technology, Harbin Institute of Technology, China
\and CVTE Research, China
\and Research Institute of Trustworthy Autonomous Systems, Southern University of Science and Technology, China \email{liuj@sustech.edu.cn}
\and Guangdong University of Technology, China
\and Yibicom Health Management Center, CVTE, China}

%
\maketitle   
\begin{abstract}
In the management of lung nodules, we are desirable to predict nodule evolution in terms of its diameter variation on Computed Tomography (CT) scans and then provide a follow-up recommendation according to the predicted result of the growing trend of the nodule. In order to improve the performance of growth trend prediction for lung nodules, it is vital to compare the changes of the same nodule in consecutive CT scans. Motivated by this, we screened out 4,666 subjects with more than two consecutive CT scans from the National Lung Screening Trial (NLST) dataset to organize a temporal dataset called NLSTt. In specific, we first detect and pair regions of interest (ROIs) covering the same nodule based on registered CT scans. After that, we predict the texture category and diameter size of the nodules through models. Last, we annotate the evolution class of each nodule according to its changes in diameter. Based on the built NLSTt dataset, we propose a siamese encoder to simultaneously exploit the discriminative features of 3D ROIs detected from consecutive CT scans. Then we novelly design a spatial-temporal mixer (STM) to leverage the interval changes of the same nodule in sequential 3D ROIs and capture spatial dependencies of nodule regions and the current 3D ROI. According to the clinical diagnosis routine, we employ hierarchical loss to pay more attention to growing nodules. The extensive experiments on our organized dataset demonstrate the advantage of our proposed method. We also conduct experiments on an in-house dataset to evaluate the clinical utility of our method by comparing it against skilled clinicians. STM code and NLSTt dataset are available at \url{https://github.com/liaw05/STMixer}.

\end{abstract}
\section{Introduction}
According to current management guidelines of lung nodules \cite{macmahon2017guidelines,wood2018lung}, it is highly desirable to perform relatively close follow-up for lung nodules detected as suspicious malignant, then give clinical guidance according to their changes during the follow-up \cite{gao2020growth}. It is recommended to extend the follow-up time for some slow-growing nodules and perform surgery on time for some nodules that are small but speculated to grow fast. However, the imprecise follow-up recommendations may yield high clinical and financial costs of missed diagnosis, late diagnosis, or unnecessary biopsy procedures resulting from false positives \cite{ardila2019end}. Therefore, it is imperative to predict the growth trend of lung nodules to assist doctors in making more accurate follow-up decisions, thereby further reducing false positives in lung cancer diagnosis. Although deep learning approaches have been the paradigm of choice for fast and robust computer-aided diagnosis of medical images \cite{ardila2019end,zhang2020learning}, there are few related studies on the use of data-driven approaches for assessing the growth trend of lung nodules, since it is hard to acquire large-scale sequential data and without ground-truth dynamic indicators.

Current clinical criteria for assessing lung nodule changes rely on visual comparison and diameter measurements from the axial slices of the consecutive computed tomography (CT) images \cite{larici2017lung}. The variation of nodule diameter between two consecutive CT scans reflects the most worrying interval change on CT follow-up screening. Applying the variation of nodule diameter to evaluate the growth trend has been a routine method \cite{huang2019prediction,tan2021prediction,yoon2020prediction,tao2022prediction}. Zhe \textit{et al.} manually calculate nineteen quantitative features (including nodule diameter) of the initial CT scans to identify the growth risk of pure ground-glass nodules (GGN) \cite{shi2019quantitative}. Xavier \textit{et al.} match the same nodule given the list of nodule candidates by computing the difference in diameter between them \cite{rafael2021re}. In this paper, inspired by existing medical research \cite{shi2019quantitative,xinyue2017analysis}, we organize a new temporal dataset of CT scans to predict the evolution of lung nodules in terms of diameter variation.

To achieve this, we screened out 4,666 subjects with more than two consecutive CT scans from the NLST dataset to organize a temporal CT dataset called NLSTt. We first detect regions of interest (ROIs) covering nodules after CT registration. Then, we pair the sequential 3D ROIs with the same nodule, followed by segmentation and classification, nodule types and diameter sizes are automatically annotated. Last, we assign one of three evolution classes (dilatation, shrinkage, stability) for each nodule in terms of its changes in diameter size. Based on automatically labeling evolution classes of each nodule in the NLSTt dataset, we further perform manually double-blind annotation and experienced review to acquire reliable labels for training a deep learning network to address the growth trend prediction. Given the inputs of 3D ROI pairs, we first introduce a siamese encoder \cite{fang2020attention,rafael2021re} to extract local features (lesion region) of two ROIs and global features of the current ROI. In order to jointly exploit the interval differences of lesions as well as the spatial dependencies between lesions and whole ROIs, we design a novel spatial-temporal mixer (STM) to leverage spatial and temporal information. In addition, we employ hierarchical loss (H-loss) \cite{yang2020hierarchical} to pay more attention to the nodule growth (dilatation class) related to possible cancerogenesis in clinical practice. 

Our \textbf{\textit{contributions}} are summarized as follows: (1) To drive the prediction of lung nodule evolution, we organize a new temporal CT dataset called NLSTt by combing automatic annotation and manual review. (2) We propose a spatial-temporal mixer (STM) to leverage both temporal and spatial information involved in the global and lesion features generated from 3D ROI pairs. (3) We conduct extensive experiments on the NLSTt dataset to evaluate the performance of our proposed method and confirm the effectiveness of our model on an in-house dataset from the perspective of clinical practice.

\section{Materials and Methods}

\subsection{NLSTt Dataset Acquisition}
\begin{figure}[!t]
\centering
\includegraphics[width=\linewidth]{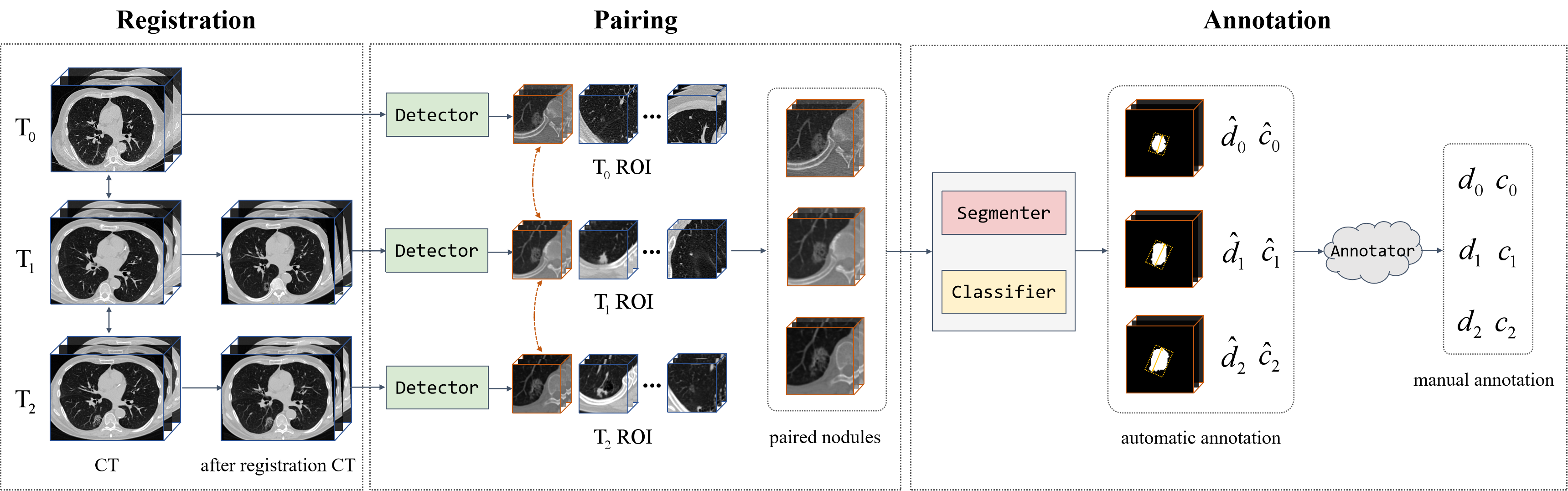} 
\caption{The pipeline of organizing the temporal CT dataset (NLSTt), including CT scan registration, ROI pairing, and class annotation. The letter $d$ denotes the diameter of lung nodules, and $c$ indicates their corresponding texture types, \textit{i.e.}, solid, part-solid (PS), ground-glass nodule (GGN).}
\label{fig:nlstt_pipeline}
\end{figure}
Lots of subjects enrolled in the NLST dataset \cite{national2011reduced} cover scans from multiple time points (typically T0, T1, and T2 scans taken one year apart) as well as biopsy confirmations for suspicious lesions \cite{veasey2020lung}. In this work, to advance the research of the growth trend prediction of lung nodules, we organize a temporal CT dataset named NLSTt by screening out 4,666 subjects from NLST, each of which has at least two CT scans up to 2 years apart from the NLST dataset.

Fig \ref{fig:nlstt_pipeline} illustrates the pipeline of our data organization approach, which aims to ascertain the evolution class and texture type of nodules in consecutive CT scans for the selected subjects. We first perform 3D image registration for the second (T1) and third (T2) CT scans in terms of the first scan (T0). After that, we identify 3D ROIs containing lung nodules by a detector, and then pair ROIs to match the same nodule in multiple 3D ROIs at different time points. Next, we employ a segmenter to automatically crop out the lesion of nodules in ROIs to calculate their diameters. At the same time, we apply a classifier to identify the texture types of nodules (\textit{i.e.}, GGN, solid, part-solid (PS)). In specific, two popular CT datasets LUNA16 \cite{setio2017validation} and LNDb \cite{pedrosa2019lndb} are used to train our models, in which the detector for ROI identification is a 3D variant of the CenterNet \cite{zhou2019objects}, the segmenter for lesion segmentation is a multi-scale 3D UNet \cite{kushnure2021ms,ronneberger2015u}, and the classifier is the variant of the segmenter attaching a fully-connected layer.

In the above pipeline of organizing the NLSTt dataset, we first make registrations for the consecutive CT scans of each patient, then detect nodules. For the detected two nodules at T0 and T1, the pair criterion is that the Euclidean distance between the center points of the two nodules is less than 1.5 millimeters. If the nodule location changes significantly between T0 and T1 ($>1.5$ millimeters), we assert that they are not the same nodule. Finally, we ask experienced clinicians to review to ensure the accuracy of paring nodules correctly. After automatically inferring the diameters and texture types, we rely on experienced clinicians to calibrate the labels by manually annotation. By combining automatic inference and manual review, we acquire the reliable label of nodules regarding the texture type and evolution class, the latter of which is determined by the diameter change of a nodule at consecutive time points. Finally, we complete the construction of the NLSTt dataset. Next, we discuss the details regarding how to construct the labels of evolution classes for nodules.

\begin{figure}[!t]
\centering
\includegraphics[width=\linewidth]{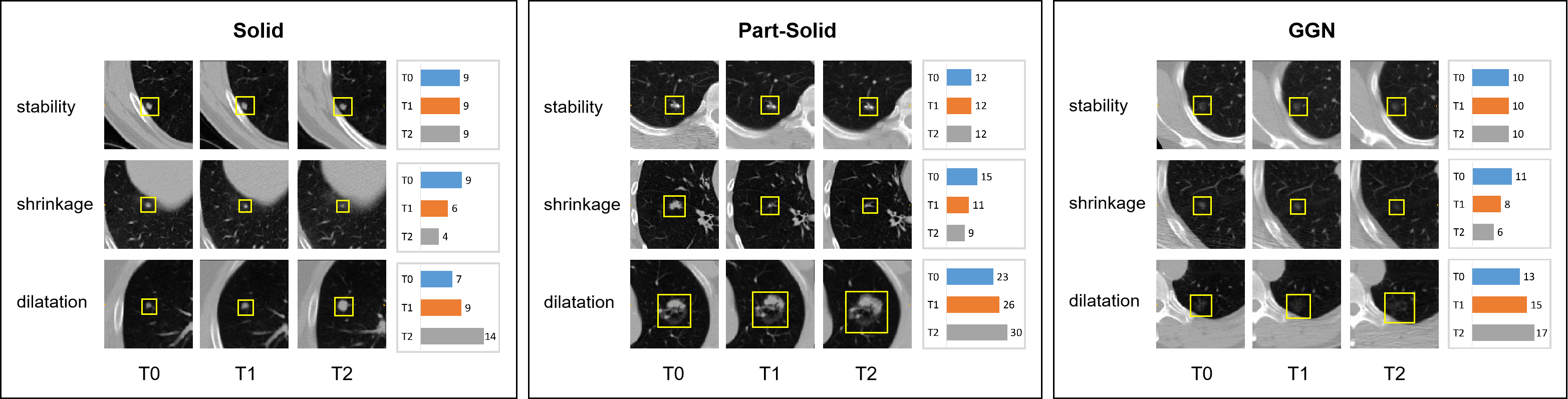} 
\caption{Three evolution classes (dilatation, shrinkage, stability) of lung nodules in the NLSTt dataset. We depict the variation of nodule diameter (in millimeter) at three consecutive time points (T0, T1, T2) for three nodule types by horizontal histograms.}
\label{fig:nlstt_sample}
\end{figure}

The evolution class of a nodule is based on the changes in diameter size. The diameter is the longest side of the smallest circumscribed rectangle on the maximal surface of the nodule mask generated by the segmenter. We formulate three evolution classes of lung nodules according to the diameter change of the same nodule at different time points as stability, dilatation, and shrinkage, as shown in Fig \ref{fig:nlstt_sample}. If the diameter variation of lung nodules at two consecutive CT scans is less than 1.5 millimeters, we refer to such trend as stability. If the diameter of lung nodules of twice CT scans varies more than 1.5 millimeters, we define the increment as dilatation and the reduction as shrinkage. In the following part, we present our proposed method for growth trend prediction.

\subsection{Spatial-Temporal Mixer}
\begin{figure}[!t]
\centering
\includegraphics[width=\linewidth]{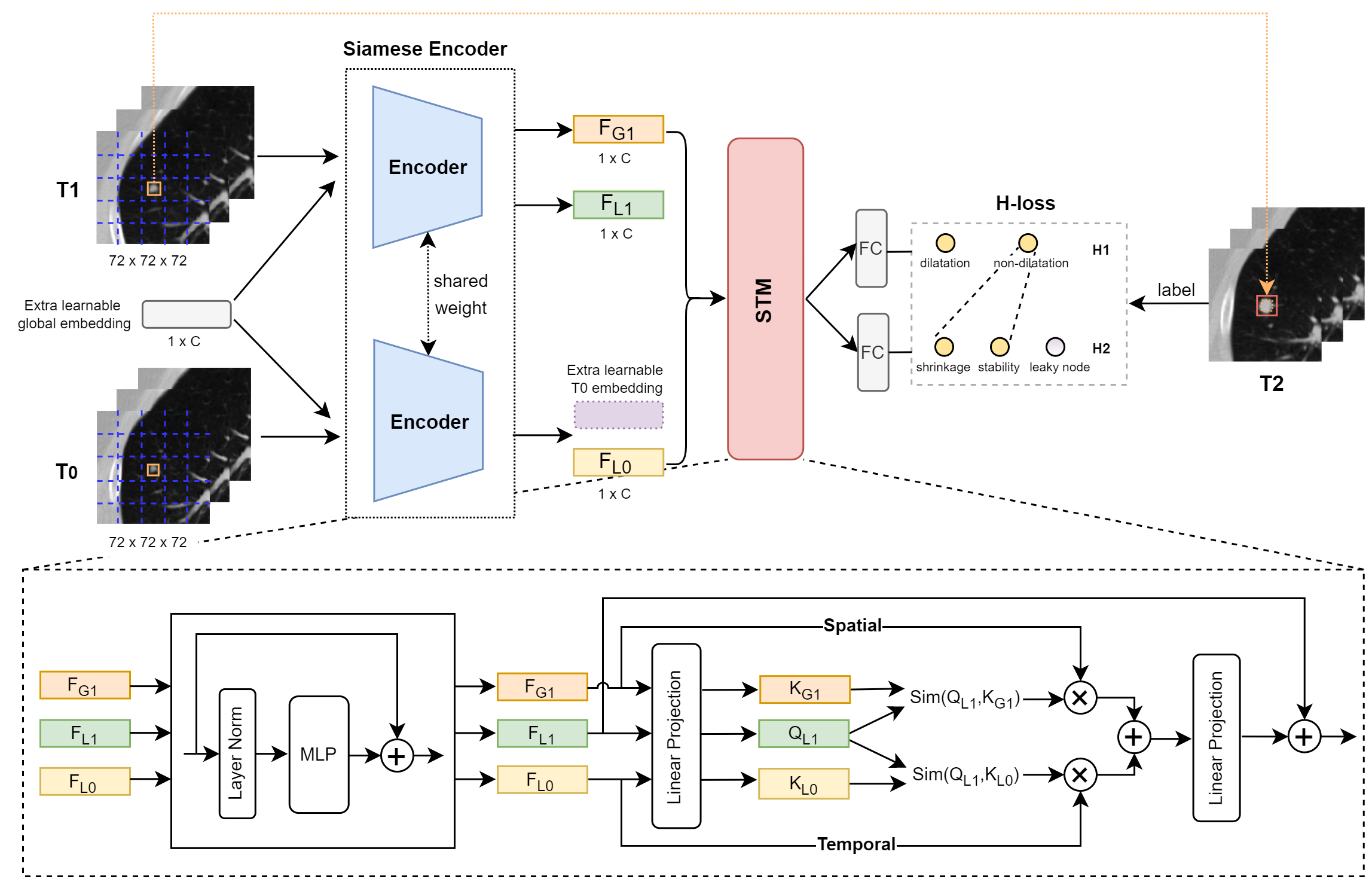} 
	\caption{Schematic of our proposed method, including a siamese encoder, a spatial-temporal mixer (STM) and a two-layer H-loss. MLP is a multilayer perception, $F_{G1}$ denotes the global information of T1, $F_{L1}$ indicates the local information of the lesion patch in T1, and $F_{L0}$ is local information of the lesion patch in T0.}
\label{fig:se_stm}
\end{figure}

Fig. \ref{fig:se_stm} overviews our proposed method for growth trend prediction, including a siamese encoder, a spatial-temporal mixer (STM), and a two-layer hierarchical loss (H-loss). For a given subject, the 3D ROI pairs (T0 and T1) containing lesions are taken from CT scans at different time points, and then fed into a siamese encoder for embedding. Besides, an extra learnable global embedding is introduced to extract the global information of 3D ROIs \cite{dosovitskiy2020image}. Both encoders in the siamese encoder share the same weights, and we adopt the vision transformer (ViT) \cite{dosovitskiy2020image,he2021masked} and convolutional neural network (CNN) \cite{he2016deep} as the backbone of the siamese encoder. We obtain three embedding vectors from sequential 3D ROIs by the siamese encoder: $F_{G1}$ contains global information of T1, $F_{L1}$ contains local information of the lesion patch in T1, and $F_{L0}$ contains local information of the lesion patch in T0. We supplement a learnable embedding $F_{L0}$ if T0 is missing, which occurs when a subject has only two CT scans. 

It is worth mentioning that the global information of ROIs is changeless on T0, T1, and T2. Hence, we only learn global information from T1 without considering the global information of T0. On the contrary, the local information of the same nodule in T0, T1, and T2 are different and highly discriminative for growth trend prediction. Therefore, we learn local information from both T0 and T1 to capture the evolving local information.

Given the embeddings obtained from the siamese encoder, we propose a spatial-temporal mixer (STM) module to leverage spatial and temporal information. We firstly introduce a layer normalization, a multi-layer perception (MLP), and a residual addition operation on the three embeddings $F_{L1}$, $F_{G1}$, and $F_{L0}$. After that, in order to fuse spatial and temporal information, we apply a linear projection to obtain a query vector $Q_{L1}$ from $F_{L1}$, and two key vectors $K_{G1}$ and $K_{L0}$ from $F_{G1}$ and $F_{L0}$, respectively. The spatial information is captured by $F_{G1} \cdot Sim(Q_{L1}, K_{G1})$, where $Sim(Q_{L1}, K_{G1})$ is the cosine similarity between the query-key pair of local and global embeddings of T1. Similarly, the temporal information is captured by $F_{L0} \cdot Sim(Q_{L1}, K_{L0})$, where $Sim(Q_{L1}, K_{L0})$ is the cosine similarity between the query-key pair of local embeddings of T1 and T0, which are collected at different time points. Next, we fuse the spatial and temporal information with an addition operation and a linear projection (LP). Finally, STM outputs an embedding based on the spatial-temporal information and the current local embeddings $F_{L1}$. 
In summary, the output of STM is computed as:
\begin{equation}
\label{equ:stm}
F_{L1} + LP(Sim(Q_{L1}, K_{G1})\cdot F_{G1} + Sim(Q_{L1}, K_{L0}) \cdot F_{L0}).
\end{equation}

\subsection{Two-layer H-loss}
After feature fusion by STM, we are ready to train a model for predicting the growth trend. We employ H-loss, which extends the weighted cross-entropy (WCE) loss function to enhance the contribution of the dilatation class by hierarchical setting. It is more cautious and attentive to the nodule growth related to possible cancerogenesis in clinical practice. Hence, it is vital to achieving high predictive accuracy for the dilatation class. To this end, we build a two-layer H-loss (H1, H2) including two separate fully-connected (FC) layer modules to analog clinical diagnosis routine. Based on three evolution classes of nodules, H1 first classifies dilatation or not for paying more attention to the dilatation class. And H2 inherits H1 to predict shrinkage or stability. The leakage node in H2 here only involves dilatation. The two-layer H-loss unifies the predictive probabilities and the ground-truth label $y$ to train our model, as follows:
\begin{equation}
\mathcal L = \alpha \cdot WCE(P_{H1}, y) + WCE(P_{H2}, y)
\label{equ:hloss},
\end{equation}
where $P_{H1}$ and $P_{H2}$ are the probability output of H1 and H2 layers, respectively. When $\alpha=0$, H-loss is a three-class WCE equivalent to H2. For H2, we set different weights for different classes in WCE to combat class imbalance. The weights of dilatation, shrinkage and stability in H2 are 1.0, 1.0, and 0.1, respectively. During model inference, we predict the evolution class of T2 by H1 or H2 layer.

\section{Experiments}

\subsection{Experimental Settings}

\begin{table}[!t]
\caption{Statistics of benchmark splits of the NLSTt dataset and in-house dataset*.}
\label{tab:nlstt_data}
\begin{center}
\begin{threeparttable}
    \begin{tabular}{|c|c|c|c|c|c|c|c|c|c|c|c|c|c|c|c|c|}
    \toprule
    \hline
    \multirow{2}{*}{\textbf{Types}} & 
    \multicolumn{4}{c|}{\textbf{Train Set}} &
    \multicolumn{4}{c|}{\textbf{Validation Set}} &
    \multicolumn{4}{c|}{\textbf{Test Set}} &
    \multicolumn{4}{c|}{\textbf{In-house Set}} \\
    \cline{2-17}
    & $\Rightarrow$ & $\Uparrow$ & $\Downarrow$ & $\sum$ 
    & $\Rightarrow$ & $\Uparrow$ & $\Downarrow$ & $\sum$ 
    & $\Rightarrow$ & $\Uparrow$ & $\Downarrow$ & $\sum$ 
    & $\Rightarrow$ & $\Uparrow$ & $\Downarrow$ & $\sum$ \\
    \hline
    \textbf{GGN} &2,496 & 153 & 34 & 2,683 & 527 & 28 & 9 & 564 & 612 & 35 & 11 & 658 & 123 & 6 & 0 & 129 \\
    \textbf{Solid} & 3,804 & 235 & 82 & 4,121 & 833 & 40 & 27 & 900 & 827 & 47 & 18 & 892 & 334 & 12 & 6 & 352 \\
    \textbf{PS} & 97 & 39 & 12 & 148 & 14 & 8 & 4 & 26 & 21 & 13 & 3 & 37 & 3 & 3 & 0 & 6 \\
    \hline
    \textbf{Totals} & 6,397 & 427 & 128 & 6,952 & 1,374 & 76 & 40 & 1,490 & 1,460 & 95 & 32 & 1,587 & 460 & 21 & 6 & 487\\
    \hline
    \bottomrule
    \end{tabular}
    \begin{tablenotes}
        \small \item[*] $\Rightarrow$ denotes stability, $\Uparrow$ indicates dilatation, $\Downarrow$ represents shrinkage, and $\sum$ aggregates the number of three evolution trends for each nodule type.
    \end{tablenotes}
\end{threeparttable}
\end{center}
\end{table}
\textbf{Datasets.} Table \ref{tab:nlstt_data} shows the statistical information of the datasets. Our organized dataset NLSTt with 4,666 subjects is split into a training set (3, 263 subjects), a validation set (701 subjects), and a test set (702 subjects). In addition, we collect CT scans of 199 subjects and adopt the same preprocessing approach used for NLSTt to organize an in-house dataset, which is used to evaluate the practicality of our model by comparing it against the clinicians.

\textbf{Optimizer.} We apply the AdamW optimizer \cite{loshchilov2016sgdr} to train our model, in which CNN encoders adopt ResNet34-3D \cite{he2016deep} and are trained from scratch, 
and ViT encoders are trained on the pre-trained model which uses the MAE method \cite{he2021masked}. The batch size is set as $B = 16$ for all the conducted methods. We warm up \cite{goyal2017accurate} the learning rate from $10e-6$ to $lr \times B / 64$ in the first 5 epochs, where $lr = 5e-4$, and then schedule the learning rate by the cosine annealing strategy \cite{loshchilov2016sgdr}. The parameters of networks are optimized in 60 epochs with a weight decay of 0.05 and a momentum of 0.9.

\subsection{Results and Discussions}

\begin{table}[!t]
\caption{AUC (in \%) of different mixers and encoders on the test and in-house sets.}
\label{tab:auc}
\begin{center}
\begin{threeparttable}
    \begin{tabular}{|c|c|c|c|c|c|}
    \toprule
    \hline
    \multirow{2}{*}{\textbf{Encoder}} & \multirow{2}{*}{\textbf{Mixer}} & \multicolumn{2}{c|}{\textbf{Test Set}} &
    \multicolumn{2}{c|}{\textbf{In-house Set}} \\
    \cline{3-6}
    & & AUC@H1 & AUC@H2& AUC@H1 & AUC@H2 \\
    \hline
    \multirow{3}{*}{\textbf{CNN}} & \textbf{Concat} & 80.8 & 75.3 & 67.2 & 67.2 \\
    & \textbf{MFC} & 81.2 & 75.2 & 69.4 & 66.7 \\
    & \textbf{LSTM} & 81.8 & 75.0 & 64.0 & 71.0 \\
    & \textbf{STM} (Ours) & 83.0 & 76.3 & \textbf{73.5} & 71.6 \\
    \hline
    \multirow{3}{*}{\textbf{ViT}} & \textbf{Concat} & 82.6 & 75.2 & 64.2 & 64.1 \\
    & \textbf{LSTM} & 82.6 & 76.3 & 67.1 & 74.7 \\
    & \textbf{STM (Ours)} & \textbf{83.6} & \textbf{77.5} & 72.8 & \textbf{78.5} \\
    \hline
    \bottomrule
    \end{tabular}
\end{threeparttable}
\end{center}
\end{table}

\textbf{Gain analysis for our STM.}
Table \ref{tab:auc} reports the Area Under the Curve (AUC) of H1 (AUC@H1) and H2 (AUC@H2) layers for four feature fusion methods (Concat, MFC, LSTM, our STM) based on CNN and ViT encoders. MFC \cite{rafael2021re} combines feature maps of 3D ROI pairs and employs CNN as the encoder. Hence, we only compare MFC and our STM based on the CNN encoder. On the test set, ViT-based STM achieves the best performances. For the in-house set, CNN-based STM obtains the best AUC@H1, and ViT-based STM gets the best AUC@H2. By further observing the high profits of ViT-based three mixers, we argue that the ViT-based siamese encoder exhibits better robustness than the CNN-based one. Furthermore, with either CNN- or ViT-based encoders, our STM brings more gains than LSTM and Concat on the test set and in-house set. Our STM also consistently outperforms MFC on two datasets in terms of two metrics. The Concat method for feature fusion only linearly combines the three embedding vectors without capturing their inter-dependencies. Hence, the predictive capability is lower than LSTM, which captures the temporal changes of two lesion features ($F_{L0}$, $F_{L1}$) extracted from T0 and T1. Besides the interval differences of lesion features at different time points. Our STM also exploits spatial dependencies of the global features $F_{G1}$ and lesion features $F_{L1}$ of T1, thus achieves the best performance.

\begin{table}[!t]
\caption{ACC and Kappa of nodule types of the test set on different methods.}
\label{tab:test_type}
\begin{center}
\begin{threeparttable}
    \begin{tabular}{|c|c|c|c|c|c|c|c|c|c|c|c|c|}
    \toprule
    \hline
    \multirow{3}{*}{\textbf{Method}} & \multicolumn{6}{c|}{\textbf{Test Set}} & \multicolumn{6}{c|}{\textbf{Extra In-house Set}} \\
    \cline{2-13}
    & \multicolumn{3}{c|}{\textbf{Accuracy}} & \multicolumn{3}{c|}{\textbf{Kappa}} & \multicolumn{3}{c|}{\textbf{Accuracy}} & \multicolumn{3}{c|}{\textbf{Kappa}} \\
    \cline{2-13}
    & GGN & Solid & PS & GGN & Solid & PS & GGN & Solid & PS & GGN & Solid & PS \\
    \hline
    \textbf{CNN+STM} & 90.9 & 88.2 & 56.8 & 27.6 & 25.7 & 26.2 & 87.6 & 91.2 & 58.1 & 14.4 & 7.6 &  27.2 \\
    \textbf{ViT+STM} & \textbf{92.4} & \textbf{91.6} & \textbf{59.5} & \textbf{29.1} & \textbf{33.7} & \textbf{29.2} & \textbf{93.8} & 90.6 & 60.5 & \textbf{46.9} & 13.4 & \textbf{29.2} \\
    \hline
    \textbf{Clinician A} & - & - & - & - & - & - & 85.3 & 93.2 & 60.5 & 19.0 & \textbf{19.8} & 23.9 \\
    \textbf{Clinician B} & - & - & - & - & - & - & 86.0 & \textbf{94.0} & \textbf{62.8} & 21.0 & 14.0 & 20.2 \\
    \hline
    \bottomrule
    \end{tabular}
\end{threeparttable}
\end{center}
\end{table}

\begin{table}[!t]
\caption{AUC of two-layer H-loss with varying $\alpha$ on CNN encoders and our STM.}
\label{tab:hloss_alpha}
\begin{center}
\begin{threeparttable}
    \begin{tabular}{|c|c|c|c|c|c|c|}
    \toprule
    \hline
    \multirow{2}{*}{$\bm{\alpha}$} & \multicolumn{3}{c|}{\textbf{Test Set}} &
    \multicolumn{3}{c|}{\textbf{In-house Set}} \\
    \cline{2-7}
    & AUC@H1 & AUC@H2 & AUC@H2-D & AUC@H1 & AUC@H2 & AUC@H2-D \\
    \hline
    \textbf{0.0} & - & 73.2 & 80.3 & - & 62.7 & 63.5  \\
    \textbf{0.5} & 82.8 & 78.2 & 82.8 & 71.1 & 71.1 & 66.6  \\
    \textbf{1.0} & 83.0 & 76.3 & \textbf{83.4} & 73.5 & 71.6 & \textbf{73.6}  \\
    \hline
    \bottomrule
    \end{tabular}
\end{threeparttable}
\end{center}
\end{table}

\textbf{Discussion of clinical practice.}
Based on the confirmation of the advantage of our STM, we further observe its performance on three nodule types and clinical applications. Since there are only six part-solid samples in the in-house set, we build an extra in-house set, which uses all the 37 part-solid samples from the test set to assemble a total of 43 part-solid samples for evaluation. As Table \ref{tab:test_type} shows, ViT-based STM obtains better Accuracy and Kappa \cite{Spitzer1967} than CNN-based STM on three nodule types of the test set. By comparing with two skilled clinicians, ViT-based STM outperforms clinicians A and B on GGN while slightly weaker on solid. For part-solid, ViT-based STM achieves better Kappa but lower Accuracy than clinicians. This demonstrates that our model can carry out clinical practice in terms of GGN, solid, and part-solid. We show several cases predicted by our model and clinicians in Appendix A1.

\textbf{Utility of two-layer H-loss.}
Table \ref{tab:hloss_alpha} shows the ablation studies of the hyper-parameter $\alpha$ in Eq. (\ref{equ:hloss}). AUC@H2-D is the AUC of the dilatation class on the H2 layer. According to AUC@H2 on the test and in-house sets, two-layer H-loss exhibits a significant advantage over the H2 layer ($\alpha=0$) alone. Among the three evolution classes, considering clinical meaning, we preferentially ensure the predictive accuracy of the growth trend. Hence, we pay more attention on the dilatation class with the help of the two-class H1 layer. The significant difference of AUC@H2-D between $\alpha=0$ and $\alpha=1$ demonstrates the benefits of our strategy.

\section{Conclusions}
In this work, we explore how to predict the growth trend of lung nodules in terms of diameter variation. We first organize a temporal CT dataset including three evolution classes by automatic inference and manual review. Then we propose a novel spatial-temporal mixer to jointly exploit spatial dependencies and temporal changes of lesion features extracted from consecutive 3D ROIs by a siamese encoder. We also employ a two-layer H-loss to pay more attention to the dilatation class according to the clinical diagnosis routine. The experiments on two real-world datasets demonstrate the effectiveness of our proposed method.

%
%
%
\bibliographystyle{splncs04}
\bibliography{tse}

\end{document}